\documentclass[final]{elsarticle}

\usepackage{amssymb,amsmath,amstext,amsthm}
  \usepackage{paralist}
  \usepackage{graphics} 
  \usepackage{epsfig} 
\usepackage{graphicx}  
\usepackage{epstopdf}
 \usepackage[colorlinks=true]{hyperref}
\usepackage{capt-of}
\usepackage{textcomp}
\usepackage{mathrsfs}
\usepackage{tabularx}

\theoremstyle{definition}

\newcommand{\ds}{\displaystyle}

\def \ld{\left\|}
\def \rd{\right\|}

\def \ld[{[\![}
\def \rd]{]\!]}

\def \I{\mathrm{I}}
\def \II{\mathrm{I\hspace{-.1em}I}}
\def \III{\mathrm{I\hspace{-.1em}I\hspace{-.1em}I}}
\def \IV{\mathrm{I\hspace{-.1em}V}}

\newcommand{\um}{\mbox{\textmu} \mathrm{m}}
\newcommand{\rhour}{\mathrm{h}}
\newcommand{\RU}{\, \mathrm{RU}}

\def\vector#1{\mbox{\boldmath$#1$}}

\newcommand{\beginsupplement}{%
        \setcounter{table}{0}
        \renewcommand{\thetable}{S\arabic{table}}%
        \renewcommand{\figurename}{Video}
        \setcounter{figure}{0}
        \renewcommand{\thefigure}{S\arabic{figure}}%
}

\begin{document}

\begin{frontmatter}

\title{
A population dynamics model of cell-cell adhesion\\ incorporating population pressure and density saturation
}


\author[carrillo]{Jose A. Carrillo}
\ead{carrillo@imperial.ac.uk}
\address[carrillo]{Department of Mathematics, Imperial College London, South Kensington Campus, London SW7 2AZ, UK}

\author[murakawa]{Hideki Murakawa\corref{cor1}}
\cortext[cor1]{Corresponding author.} 
\ead{murakawa@math.kyushu-u.ac.jp, hideki.murakawa@gmail.com}
\address[murakawa]{Faculty of Mathematics, Kyushu University, 744 Motooka, Nishiku, Fukuoka, 819-0395, Japan}

\author[sato]{Makoto Sato}
\ead{makotos@staff.kanazawa-u.ac.jp}
\address[sato]{Laboratory of Developmental Neurobiology, Graduate School of Medical Sciences, Mathematical Neuroscience Unit, Institute for Frontier Science Initiative, Kanazawa University, 13-1 Takaramachi, Kanazawa, Ishikawa 920-8640, Japan}

\author[togashi]{Hideru Togashi}
\ead{htogashi@med.kobe-u.ac.jp}
\address[togashi]{Division of Molecular and Cellular Biology, Department of Biochemistry and Molecular Biology, Kobe University Graduate School of Medicine, 7-5-1, Kusunoki-cho, Chuo-ku, Kobe 650-0017, Japan} 

\author[trush]{Olena Trush}
\ead{trush@stu.kanazawa-u.ac.jp, olena971@gmail.com}
\address[trush]{Laboratory of Developmental Neurobiology, Graduate School of Medical Sciences, Kanazawa University, 
13-1 Takaramachi, Kanazawa, Ishikawa 920-8640, Japan}

\begin{abstract}
We discuss several continuum cell-cell adhesion models based on the underlying microscopic assumptions. We propose an improvement on these models leading to sharp fronts and intermingling invasion fronts between different cell type populations. The model is based on basic principles of localized repulsion and nonlocal attraction due to adhesion forces at the microscopic level. The new model is able to capture both qualitatively and quantitatively experiments by Katsunuma et al. (2016) [J. Cell Biol. 212(5), pp. 561--575]. We also review some of the applications of these models in other areas of tissue growth in developmental biology. We finally explore the resulting qualitative behavior due to cell-cell repulsion.
\end{abstract}

\end{frontmatter}


\section{Introduction}

More than 50 years ago, Steinberg proposed the differential adhesion hypothesis to explain the force directing cell motility during morphogenesis \cite{steinberg1,steinberg2,steinberg3,steinberg4}. 
Selective cell-cell adhesion between different cell types is fundamental for sorting different cell types in morphogenesis. For examples, cells dissociated from various tissues of vertebrate embryos preferentially reaggregate with cells from the same tissue when they are mixed and cultured together. This process is mediated by various cell adhesion molecules that hold cells together by homophilic or heterophilic interactions between adjacent cells. The major cell adhesion molecules in vertebrates are cadherins and nectins. Cadherins are homophilic adhesion molecules involved in various morphogenetic processes. In contrast to cadherins, nectins prefer heterophilic to homophilic adhesion. 

This paper considers mathematical models capable of replicating cell sorting phenomena due to differential adhesion focusing on the different adhesive properties of molecules such as cadherins or nectins. 
A number of single-cell-based models have been developed and used for cell sorting phenomena, e.g., the cellular Potts models \cite{cgia,gg1992,potts,kapkgmh,man}, lattice-free models \cite{po,scp,HBTaylor}, the vertex dynamics models \cite{AGS,HNO,HOHK} and so on. 
These models can show excellent agreement between numerical experiments and some cell sorting phenomena. 
However, these models have some drawbacks; 
it is necessary to set many rules and parameters for numerical calculations; 
multiple factors having causal relations such as adhesion and motility have to be stipulated separately;
there are restrictions on the size and shape of cells, and it is difficult to consider aggregations of cells with complicated shapes such as neurons; 
it is very difficult to elucidate the essence of the phenomenon from analytical points of view.  
In order to overcome these drawbacks, cell population dynamics models are considered. 
Our goal in this paper is to propose a new population model to explain cell sorting due to differential adhesion.

Let us consider that we have two type of cells expressing different levels of protein surface ligands such as cadherins or nectins. Let us consider that $\{ x_i \}$, $i=1,\dots,N$, and $\{ y_j \}$, $j=1,\dots,M$ represent the positions of the nuclei of each of the cells forming these two groups and avoid dealing with cell membrane remodelling since we will work at the population level. Our assumption, introduced in \cite{CCS}, is that cells will interact with other cells either by attraction at medium distances through the formation of protusions or filopodia or by strong repulsion if the interparticle cell distance becomes very small due to volume size constraints around the nuclei. For simplicity, we assume that $N=M$ from now on. Let us also model forces exerted by cell type $j$ onto cell type $i$ as radial nonlinear springs in the direction of the centers of the nuclei, and therefore they derived from a radial potential denoted by $W^N_{ij}$, $i,j=1,2$. The basic agent based model for these two cell types reads as
\begin{align*}
	\dot x_i &= - \frac{m_1}N \sum_{j\neq i}\nabla W^N_{11}(x_i-x_j) - \frac{m_2}N\sum_{j\neq i} \nabla W^N_{12}(x_i-y_j),\\
	\dot y_i &= - \frac{m_2}N \sum_{j\neq i} \nabla W^N_{22}(y_i-y_j) - \frac{m_1}N\sum_{j\neq i} \nabla W^N_{21}(y_i-x_j).
\end{align*}
As we deal with a large number of cells, we will work at the population level, where we model cells densities rather that cell positions. With this aim, we now introduce the empirical measures associated to the above agent-based models, and describe the two populations via the mean-field scaling, the factor $1/N$, on each of these potentials with equal number of agents in order to keep a finite mass $m_i$, $i=1,2$, for each of the populations. We denote by $\rho_i$, $i=1,2$, the populations densities of each cell-type with total masses $m_i$, $i=1,2$, that in the limit of large number of agents $N\to \infty$, are given by the limits of the empirical measures, i.e.
$$
\rho_1(x,t) \simeq \frac{m_1}N \sum_{i=1}^N \delta_{x_i(t)} \qquad \mbox{and} \qquad \rho_2(x,t) \simeq  \frac{m_2}N \sum_{i=1}^N \delta_{y_i(t)} \qquad \mbox{ as $N\to \infty$.}
$$
Here, the notation $\delta_{x_o}$ refers to the Dirac Delta measure at the point $x_o$. Let us remark that agent-based models of this type with a finite number of cells are also interesting in detailed models where differential adhesion is important, see \cite{CCS} and the references therein. They include rosette formation in the early migration of the zebrafish lateral line primordium \cite{pri1,pri2}, whose dynamics are fundamental for the correct embryonic development of the animal, and zebrafish stripe skin patterning \cite{VS}.

We now put further assumptions on the scaling of these potentials reflecting the attraction for distances less than some cut-off radius $R$ together with the volume size restriction modelled by localized repulsion \cite{CC}. Each potential is scaled in $N$ in such a way that $W^N_{ij}\simeq \epsilon \delta_0 + W_{ij}$ as $N\to \infty$ with $W_{ij}$ being radially symmetric, compactly supported on the ball of radius $R$, and strictly attractive inside this ball. The limit $N\to\infty$ leads to the following system for the densities $\rho_i$, $i=1,2$,
\begin{equation}
\label{model_chs}
\left\{
\begin{aligned}
    \partial_t \rho_1 &= \nabla \cdot\Big(\rho_1 \nabla \big(W_{11}\star\rho_1 + W_{12}\star\rho_2 + \epsilon (\rho_1+\rho_2)\big)\Big),\\
    \partial_t \rho_2 &= \nabla \cdot\Big(\rho_2 \nabla \big(W_{22}\star\rho_2 + W_{21}\star\rho_1 + \epsilon (\rho_1+\rho_2)\big)\Big).
\end{aligned}
\right.
\end{equation}
Here, star $\star$ denotes the convolution of two functions. 
The rigorous derivation for one single cell type can be done following the blueprint in \cite{Oe}, see \cite{BCM,BV} and the references therein too. This basic model for two populations shows very rich dynamical properties and complex set of stationary states and stability \cite{BFH,CCH,CHS,BDFS}. However, it does not establish any upper bound on the maximal density of cells, density saturation, that is sensible for cell population models and it does not include reaction terms to take into account cell apoptosis and cell division/growth. Another well-known model was proposed by Armstrong, Painter and Sherratt~\cite{aps2006}. They include interaction and reaction terms to model cell migration and growth/death with nonlinearities depending on the two cell densities $u,v$ while assuming linear diffusion for both, it reads as
\begin{equation}
\label{model_aps}
\left\{
\begin{aligned}
&\frac{\partial u}{\partial t} = \Delta u -\nabla \cdot \left(u \vector{K}_{\! g1}(u,v)\right)+f_1(u,v),\\
&\frac{\partial v}{\partial t} = \Delta v-\nabla \cdot \left(v \vector{K}_{\! g2}(u,v)\right)+f_2(u,v). 
\end{aligned}
\right.
\end{equation}
Here, the advection velocities $\vector{K}_{\! gi}(u,v)$, $i=1,2$, are nonlocal terms modelling the cell adhesion with neighbouring cells with a cut-off radius $R$ but proportional not to the density of cells but to a suitable nonlinear saturation of the density to impose density saturation. We will explain this in full details below. The resulting model is able to capture concentrated densities but it does not lead to full segregation and/or sharp boundaries \cite{BDM,CHS} between cell types as in the case of \eqref{model_chs} due to the linear diffusion terms and the lack of population pressure, similar drawbacks are shared by classical segregation models in population dynamics \cite{SKT,Shi80} and references therein. Similar systems to \eqref{model_aps} without cross-diffusion as in \eqref{model_chs} appear in modelling cell adhesion in other mathematical biology contexts such as zebrafish patterning and tumour growth models, see \cite{GC,DTGC,PBSG,VS} for instance.

A variation of the model in \eqref{model_aps} taking into account the population pressure coming from volume size instead of linear diffusion, as in \eqref{model_chs},  was proposed by part of the authors in \cite{mt}. They obtained sharp boundaries and segregated densities but several drawbacks due to the modelling of the migration term has been identified since then. We will elaborate on this in section 2. In this work, we propose a variation of the models in \cite{aps2006,mt} taking into account the population pressure coming from volume size instead of linear diffusion as in \eqref{model_chs} together with a density saturation mechanism different from \eqref{model_aps} for which forces are still computed linearly with respect to the local population but the saturation of the density is taken care on the mobility term. We will present the model and its improvements with respect to the previous literature in Section 2.
We here validate that this model captures well the differential adhesion hypothesis \cite{steinberg1,steinberg2,steinberg3,steinberg4} explaining patterns of cell sorting and cellular movement during morphogenesis with thermodynamic principles, see a differential cell adhesion scheme in Fig.~\ref{fig_DAH}A below.


We will show in Section 3 that this model is able to capture qualitatively the cell sorting mechanism in \cite{katsunuma}. 
Moreover, once the parameters have been carefully identified, the model is able to capture quantitatively finite interpenetration zone lengths between cell types and their invasion fronts for the experiments developed in  \cite{katsunuma}. 
The continuous models have some advantages to deal with problems that were difficult to be handled with conventional computable models such as single-cell-based models introduced above. 
Indeed, the continuous models have been applied  to real problems in life sciences recently. 
Two applications will be reviewed briefly in Section~4. 
Our model can represent not only cell-cell adhesion but also cell-cell repulsion. 
In Section~5, we show the advantage by carrying out numerical experiments of the single-component model.


\section{Comparison and derivation of cell sorting models}

Let us start by introducing the Murakawa--Togashi model~\cite{mt} for cell-cell adhesion. We first deal with a single population of cells for the sake of simplicity. 
The population density at position $x$ and time $t$ is denoted by $u(x,t)$. The model is based on the following conservation of mass:
\[
\frac{\partial u}{\partial t}=-\nabla \cdot (u (\vector{V}_{\!\! \mathrm{p}}+\vector{V}_{\!\! \mathrm{a}})), 
\]
where the velocity vector is composed of velocities due to pressure $\vector{V}_{\!\! \mathrm{p}}$ and adhesion $\vector{V}_{\!\! \mathrm{a}}$. 
Assume that the pressure $p$ is proportional to the population density, namely, 
\begin{equation}
\label{pressure_density}
\vector{V}_{\!\! \mathrm{p}}=-\nabla p = -c_\mathrm{p} \nabla u. 
\end{equation}
Here $c_\mathrm{p}$ is called the dispersivity. More generally, the pressure can also be chosen as a nonlinear function of density, namely, $p=\chi(u)$. 
Following the argument by Armstrong, Painter and Sherratt~\cite{aps2006}, the adhesion velocity vector is given as follows: 
\begin{align}
\vector{V}_{\!\! \mathrm{a}}(\vector{x})=\frac{\phi}{R}\vector{K}_{\! g}(u)(\vector{x})
=\frac{\phi}{R}\int_0^R\!\!\!  \int_{S^{d-1}} a\, g(u(\vector{x}+r\vector{\eta}))\, \omega(r)r^{d-1} \vector{\eta} \, d\vector{\eta} dr, \label{velvec_original}
\end{align}
where $\phi$ is a constant of proportionality related to viscosity, $R$ is a positive constant called sensing radius, $a$ is an adhesive strength parameter, 
$S^{d-1}$ is the $d$-dimensional unit spherical surface. Note that in one dimension $S^0$ contains only two points and the integral becomes a sum. 
The function $g$ represents how the adhesive force depends on the local cell density, which is defined 
\[
g(u)
= 
\left\{
\begin{array}{lll}
u(1-u/m) & \mathrm{if} \ \ u<m,\\
0 & \mathrm{otherwise}, 
\end{array}
\right.
\]
where $m$ denotes the crowding capacity of the population. The function $\omega$ describes how the force is dependent on the distance from $\vector{x}$. In fact, by 
defining $\Omega(\vector{x})=\Omega (|\vector{x}|)$, then $\nabla \Omega(\vector{x})=\omega(r) \frac{\vector{x}}{|\vector{x}|}$. Thus, the adhesion velocity can also be written as
\begin{align*}
\vector{V}_{\!\! \mathrm{a}}(\vector{x})=\frac{\phi}{R}\int_{B(0,R)}\!\!\! a\, g(u(\vector{x}+\vector{y}))\, \nabla \Omega(\vector{y}) \, d\vector{y} =
-\frac{\phi}{R}\int_{B(\vector{x},R)}\!\!\! a\, g(u(\vector{y}))\, \nabla \Omega(\vector{x-y}) \, d\vector{y}, 
\end{align*}
that is similar to the aggregation equation models studied in \cite{MEK99,TBL06,BCL,CDFLS11} except for the nonlinearity $g(u)$ in the convolution. We will comment of the choice of the force $\omega(r)$ below.
Adding the logistic growth term, the following model is obtained:
\begin{equation*}
\frac{\partial u}{\partial t}=c_\mathrm{p}\nabla \cdot (u\nabla u) - \frac{\phi}{R} \nabla \cdot (u\vector{K}_{\! g}(u))+ bu(1-u/k). 
\end{equation*}
Here, $b$ represents the growth rate and $k$ is the carrying capacity. We now rescale the system to produce a dimensionless formulation, with this goal we choose the following units 
\begin{equation}
\label{rescale}
\begin{array}{cc}
\ds x^*=\frac{x}{R},\quad 
t^*=\frac{c_\mathrm{p} m}{R^2}t,\quad
u^*=\frac{u}{m}, \quad
\ds a^*=\frac{R^d \phi}{c_\mathrm{p}}a,\quad
b^*=\frac{R^2 b}{c_\mathrm{p} m},\quad
k^*=\frac{k}{m}, 
\end{array}
\end{equation}
and dropping the stars, the following non-dimensional model for the single population of cells is obtained. 
\begin{equation}
\label{eq_original_1c}
\frac{\partial u}{\partial t}=\nabla \cdot (u\nabla u) -  \nabla \cdot (u\vector{K}_{\! g}(u))+ f(u). 
\end{equation}
Here, the functions $\vector{K}_{\! g}$ and $f$ are redefined as follows:
\begin{equation}
\vector{K}_{\! g}(u)(\vector{x})=a\,\int_0^1\!\!\!  \int_{S^{d-1}} g(u(\vector{x}+r\vector{\eta}))\, \omega(r)r^{d-1} \vector{\eta} \, d\vector{\eta} dr, \label{def_K1}
\end{equation}
\begin{equation}
\label{def_g}
g(u)
= 
\left\{
\begin{array}{lll}
u(1-u) & \mathrm{if} \ \ u<1,\\
0 & \mathrm{otherwise}, 
\end{array}
\right.
\end{equation}
\begin{equation*}
\label{def_f}
f(u) = bu\left(1-u/k\right),
\end{equation*}
reducing the model to 3 parameters $(a,b,k)$: adhesion strength, reaction strength and dimensionless carrying capacity.
This derivation can be extended to a model of two types of cell populations by introducing a total population pressure leading to
\begin{equation}
\label{model_original}
\left\{
\begin{aligned}
&\frac{\partial u}{\partial t} = \nabla \cdot \left(u \nabla(u+v)\right)-\nabla \cdot \left(u \vector{K}_{\! g1}(u,v)\right)+f_1(u,v),\\
&\frac{\partial v}{\partial t} = \nabla \cdot \left(v \nabla(u+v)\right)-\nabla \cdot \left(v \vector{K}_{\! g2}(u,v)\right)+f_2(u,v). 
\end{aligned}
\right.
\end{equation}
Here the functions $\vector{K}_{\! gi}$ ($i=1,2$) are given by
\begin{equation}
\label{def_Kg}
\begin{aligned}
\vector{K}_{\! gi}(u,v)(\vector{x})=
\int_0^1\!\!\! \int_{S^{d-1}} &\Big[ a_{i1}g_{i1}(u(\vector{x}+r\vector{\eta}),v(\vector{x}+r\vector{\eta})) \\
&+
a_{i2}g_{i2}(u(\vector{x}+r\vector{\eta}),v(\vector{x}+r\vector{\eta}))
\Big]
 \omega(r)r^{d-1}\vector{\eta}\, d\vector{\eta} dr, 
\end{aligned}
\end{equation}
where $a_{ij}$ ($i,j=1,2$) denote rescaled adhesive strength parameters between the $i{\rm th}$ and $j{\rm th}$ cell populations, and the functions $g_{ij}$ are defined by
\[
\begin{aligned}
g_{11}(u,v) &= g_{21}(u,v) = 
\left\{
\begin{array}{lll}
u(1-u-v) & \mathrm{if} \ \ u+v<1,\\
0 & \mathrm{otherwise},
\end{array}
\right.
\end{aligned}
\]
\[
\begin{aligned}
g_{22}(u,v) &= g_{12}(u,v) = 
\left\{
\begin{array}{lll}
v(1-u-v) & \mathrm{if} \ \ u+v<1,\\
0 & \mathrm{otherwise}. 
\end{array}
\right.
\end{aligned}
\]
The growth terms $f_1$ and $f_2$ in \eqref{model_original} are defined by 
\begin{equation}
\label{logistic_growth}
\begin{aligned}
f_1(u,v) = b_1u\left(1-(u+v)/k_1\right),\\
f_2(u,v) = b_2v\left(1-(u+v)/k_2\right),
\end{aligned}
\end{equation}
where $b_i$ and $k_i$ are the rescaled growth rate and the ratio of the carrying capacity of the $i$th population to the crowding capacity, respectively. 

\begin{figure}[!ht]
\centering
\includegraphics[width=0.8\textwidth,clip]{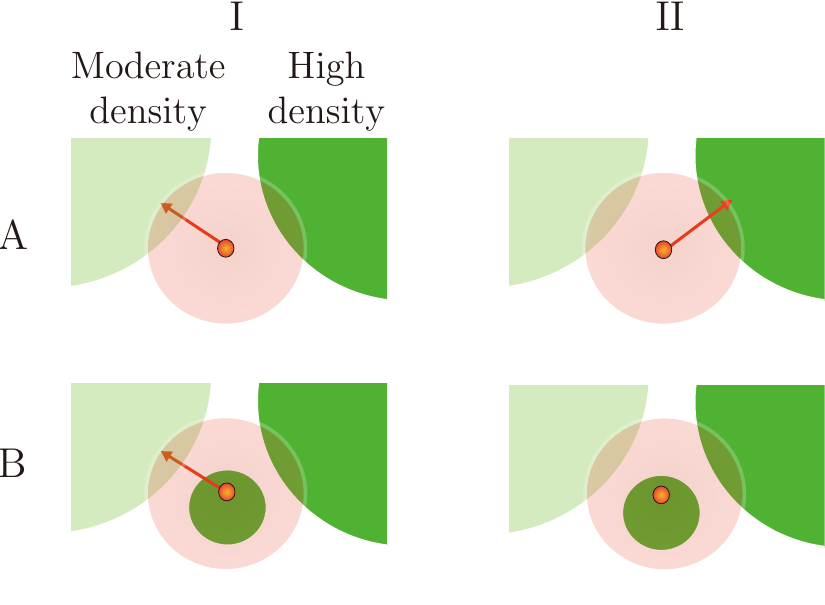}
\caption{
Schematic illustrations of cell movement due to adhesive forces towards red cell. ($\I$) Adhesive forces due to \eqref{velvec_original}, ($\II$) idealistic adhesive forces. Green regions illustrate cell colonies, and color intensity implies higher density of cells in the colony. The pink disk illustrates the sensing region of the red cell. 
}
\label{fig_explanation}
\end{figure}

We now introduced a new modified model with respect to \cite{mt} to improve several of its drawbacks. Let us consider two typical situations shown in Fig.~\ref{fig_explanation}, where there are two cell colonies which intersect with the sensing region of a red cell, one with moderate density and the other density is almost reaching crowding capacity. 

As shown in Fig.~\ref{fig_explanation}(I--A), according to the adhesive velocity \eqref{velvec_original}, an isolated red cell moves towards the moderate density colony because the advection term $\vector{K}_{\! gi}(u,v)(\vector{x})$ in \eqref{def_K1} penalizes more the high density region. However, the probability of finding other cells is higher in the right colony than in the left colony. 
Therefore, it is more reasonable that the force is directed towards the higher density region as shown in Fig.~\ref{fig_explanation}($\II$--A).  Another unreasonable situation is depicted in Fig.~\ref{fig_explanation}(I--B). Here, the adhesive force due to \eqref{velvec_original} is directed towards the moderate density colony even if the red cell is surrounded by highly concentrated green cell density since both small and high densities are penalized due to the form of $g(u)$ in \eqref{def_g}. However, it is more reasonable that the red cell does not move in such a situation as shown in Fig.~\ref{fig_explanation}($\II$--B). 

\begin{figure}[!ht]
\centering
\includegraphics[width=0.8\textwidth,clip]{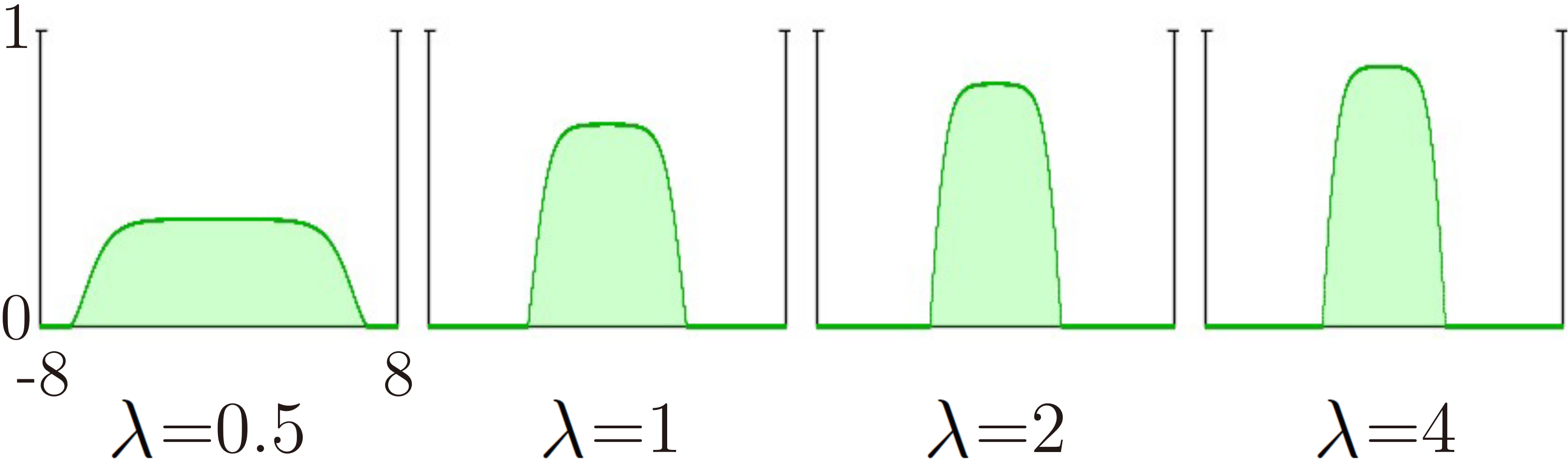}
\caption{
Steady states of \eqref{eq_outside_1c} with $f\equiv 0$ in one space dimension with
different potentials. The adhesive strength is $a=2$, the total mass is
four and the potential is $\Omega(\vector{x})=|\vector{x}|^\lambda$.
}
\label{fig_diff_potential}
\end{figure}

To realize these ideal situations, we propose to redefine the following adhesion velocity vector instead of \eqref{velvec_original}: 
\begin{align}
\vector{V}_{\!\! \mathrm{a}}(\vector{x}) 
&= a\frac{\phi}{R}(1-u/m) \int_0^R\!\!\!  \int_{S^{d-1}}  u(\vector{x}+r\vector{\eta})\, \omega(r)r^{d-1} \vector{\eta} \, d\vector{\eta} dr. \label{velvec_outside}
\end{align}
Here, each cell counts its surrounding linearly increasing in density to determine the direction of movement. The magnitude of the total force decreases as the density locally at the cell position increases.  Using the same velocity vector due to pressure population and analogous growth term as above, and applying the same rescaling as in \eqref{rescale}, we have the following modified model for single cell population: 
\begin{equation}
\label{eq_outside_1c}
\frac{\partial u}{\partial t}=\nabla \cdot (u\nabla u) - \nabla \cdot (u(1-u)\vector{K}(u))+ f(u),  
\end{equation}
where 
\begin{align*}
\vector{K}(u)(\vector{x}) 
=  a \int_0^1\!\!\!  \int_{S^{d-1}} u(\vector{x}+r\vector{\eta})\, \omega(r)r^{d-1} \vector{\eta} \, d\vector{\eta} dr . 
\end{align*}
The choice of the radial force $\omega$ can be used as a parameter too. However, the balance between the volume exclusion term and the aggregation term leads to stationary states in the absence of the reaction term $f\equiv 0$. In Fig.~\ref{fig_diff_potential}, we observe these steady states for different choices of the potential $\Omega$. 
Since their qualitative behavior is analogous for all radial attractive power-law potentials, being streched in the vertical direction for larger exponents $\lambda$, we decided to fix $\omega \equiv 1$ or equivalently $\lambda=1$ for all numerical simulations below.

All two dimensional numerical simulations in this paper are carried out in a fixed domain $[-5,5)^2$ with the periodic boundary condition. The standard  explicit upwind finite volume method is employed and the nonlocal terms are calculated by numerical integrations (see \cite{mt} for more details). 

\begin{figure}[!t]
\centering
\includegraphics[width=0.99\textwidth,clip]{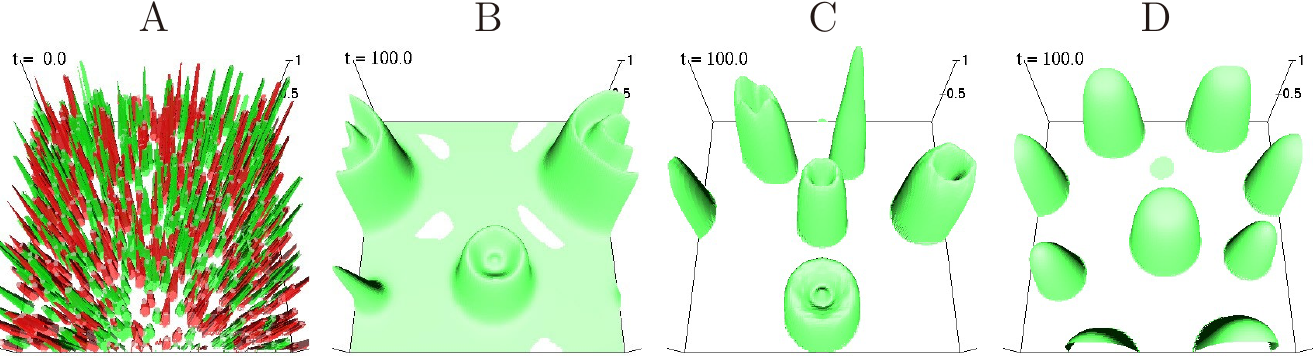}
\caption{
Numerical experiments of \eqref{model_aps}, \eqref{eq_original_1c} and \eqref{eq_outside_1c} in two space dimensions. (A) Initial distribution for the two-component numerical experiments in Fig.~\ref{fig_DAH}. Green+red implies initial distribution for the one-component simulation (B)--(D). (B) Numerical result of \eqref{model_aps} with $a=60$. (C) Numerical result of \eqref{eq_original_1c} with $a=60$.  (C) Numerical result of \eqref{eq_outside_1c} with $a=4$.}
\label{fig_1c}
\end{figure}

Let us now compare two dimensional numerical solutions of \eqref{model_aps} (Fig.~\ref{fig_1c} (B)), \eqref{eq_original_1c} (Fig.~\ref{fig_1c} (C)), and of \eqref{eq_outside_1c} (Fig.~\ref{fig_1c} (D)) without growth terms. We observe how the new interaction mechanism leads to the formation of separated smooth bumps leading to stationary states of the same shape but with different masses in contrast with the previous models leading to corrugated surfaces depending on the mass of each of the bumps.

\begin{figure}[!th]
\centering
\includegraphics[width=0.99\textwidth,clip]{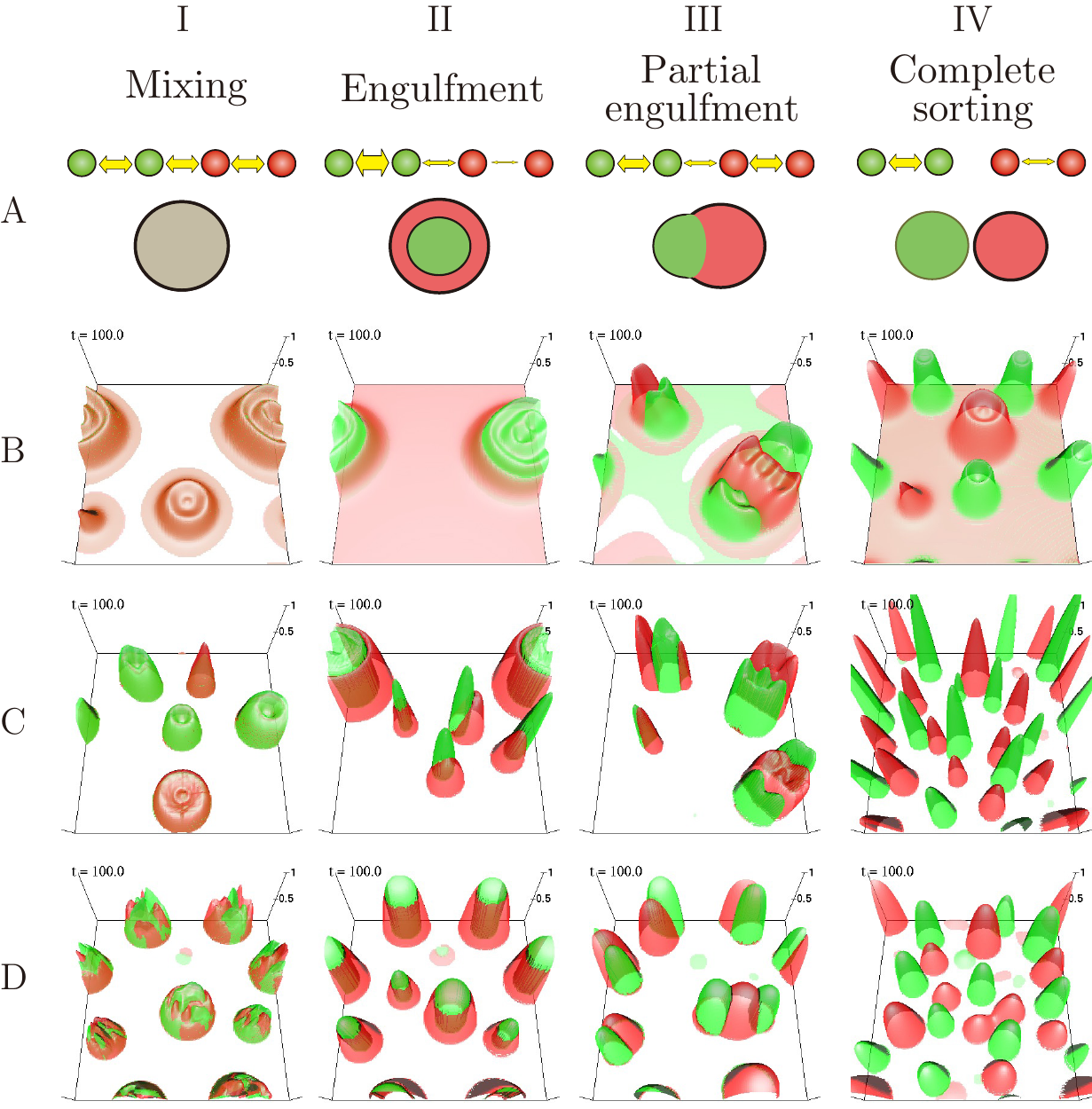}
\caption{
Numerical results for the differential adhesion hypothesis in two space dimension. (A) Schematic of the relative adhesion strengths and possible configurations. (B) Results of \eqref{model_aps}, (C) results of \eqref{model_original}, (D) results of \eqref{model_outside}. Adhesive strength parameters are as follows: 
(BC--$\I$) $a_{11}=60$, $a_{12}=a_{21}=60$, $a_{22}=60$, 
(BC--$\II$) $a_{11}=120$, $a_{12}=a_{21}=30$, $a_{22}=15$, 
(BC--$\II$) $a_{11}=60$, $a_{12}=a_{21}=30$, $a_{22}=60$, 
(BC--$\IV$) $a_{11}=60$, $a_{12}=a_{21}=0$, $a_{22}=30$, 
(D--$\I$) $a_{11}=4$, $a_{12}=a_{21}=4$, $a_{22}=4$, 
(D--$\II$) $a_{11}=6$, $a_{12}=a_{21}=4$, $a_{22}=2$, 
(D--$\III$) $a_{11}=6$, $a_{12}=a_{21}=2$, $a_{22}=4$, 
(D--$\IV$) $a_{11}=6$, $a_{12}=a_{21}=0$, $a_{22}=4$. 
}
\label{fig_DAH}
\end{figure}

Extending this idea to two-component problem leads to the system 
\begin{equation}
\label{model_outside}
\left\{
\begin{aligned}
&\frac{\partial u}{\partial t} = \nabla \cdot \left(u \nabla(u+v)\right)-\nabla \cdot \left(u (1-u-v) \vector{K}_{\! 1}(u,v)\right)+f_1(u,v),\\
&\frac{\partial v}{\partial t} = \nabla \cdot \left(v \nabla(u+v)\right)-\nabla \cdot \left(v (1-u-v)\vector{K}_{\! 2}(u,v)\right)+f_2(u,v). 
\end{aligned}
\right.
\end{equation}
Here the functions $f_i$ ($i=1,2$) are defined as in \eqref{logistic_growth}, and $\vector{K}_{\! i}$ ($i=1,2$) are the functions $\vector{K}_{\! gi}$ of \eqref{def_Kg} with $g_{i1}(u,v)=u$ and $g_{i2}(u,v)=v$, namely, 
\begin{equation*}
\label{def_K}
\begin{aligned}
\vector{K}_{\! i}(u,v)(\vector{x})=
\int_0^1\!\!\! \int_{S^{d-1}} &\Big[ a_{i1}u(\vector{x}+r\vector{\eta}) 
+
a_{i2}v(\vector{x}+r\vector{\eta})
\Big]
 \omega(r)r^{d-1}\vector{\eta}\, d\vector{\eta} dr, 
\end{aligned}
\end{equation*}
Now let us look at two dimensional numerical simulations of  \eqref{model_aps}, \eqref{model_original} and \eqref{model_outside} for the differential adhesion hypothesis (Fig.~\ref{fig_DAH}). Here, we set $f_i\equiv 0$ ($i=1,2$).  We use translucent green (resp. translucent red) to plot $u$ (resp. $v$) in order to make it easy to understand whether the two types of cells are mixed or separated. The region where $u,v<10^{-3}$ is painted in white. 
Model~\eqref{model_aps} does not lead to segregation and the two types of cells are intermixed in engulfment (Fig.~\ref{fig_DAH}(B--$\II$)), partial engulfment (Fig.~\ref{fig_DAH}(B--$\III$)) and complete sorting (Fig.~\ref{fig_DAH}(B--$\IV$)) cases because of an oversimplification of the diffusion part of the model. 
Both models \eqref{model_original} (Fig.~\ref{fig_DAH}(C)) and \eqref{model_outside} (Fig.~\ref{fig_DAH}(D)) can replicate different cell sorting patterns explained by the differential adhesion hypothesis and show sharp boundaries between the green and red cells. 
As seen in one-component numerical simulations, 
Model~\eqref{model_outside} leads to smooth bumps, whereas Model~\eqref{model_original} leads to corrugated surfaces. 
For the mixing pattern Fig.~\ref{fig_DAH}(I), 
Model~\eqref{model_aps} and \eqref{model_original} show that the green and red cells are intermixed about fifty-fifty. In contrast, in Model~\eqref{model_original}, these cells form marble pattern colonies. 
Since the green and red cells are expressing the same type of nectin molecule in these cases and these cells are scattered apart in the initial state, the state of being mixed randomly like Fig.~\ref{fig_DAH}($\I$--D) is more realistic. 
Incidentally, the numerical solutions $u$(green)+$v$(red) in Fig.~\ref{fig_DAH} ($\I$--B), ($\I$--C) and ($\I$--D) coincide with those in Fig.~\ref{fig_1c} (B), (C) and (D), respectively. 
Thus, the colonies in Fig.~\ref{fig_DAH}($\I$--D) are smooth.  

In this paper, we suggested specific choices of the population pressure in \eqref{pressure_density} and the density saturation in \eqref{velvec_outside}. 
Of course, one can consider more general form of the model as follows: 
\begin{equation*}
\left\{
\begin{aligned}
&\frac{\partial u}{\partial t} = \nabla \cdot \left(u \nabla \chi_1(u,v)\right)-\nabla \cdot \left(u g_1(u,v) \vector{K}_{\! 1}(u,v)\right)+f_1(u,v),\\
&\frac{\partial v}{\partial t} = \nabla \cdot \left(v \nabla \chi_2(u,v)\right)-\nabla \cdot \left(v g_2(u,v) \vector{K}_{\! 2}(u,v)\right)+f_2(u,v). 
\end{aligned}
\right.
\end{equation*}
Here, $\chi_i$ and $g_i$ ($i=1,2$) are given functions.

\section{Numerical simulations for the Togashi et al.'s experiment}
\label{sec_TE}

Nectins are immunoglobulin-like cell-adhesion molecules that compose a family of four members (nectin-1, -2, -3, and -4)~\cite{tn}. Nectins prefer heterophilic interactions to homophilic ones, in contrast with the homophilic nature of cadherin interactions. For example, when cells expressing nectin-1 and -3 are mixed, these cells are arranged in a mosaic pattern due to their heterophilic interactions~\cite{togashi2,togashi1}. 

\begin{figure}[!ht]
\centering
\includegraphics[width=0.9\textwidth,clip]{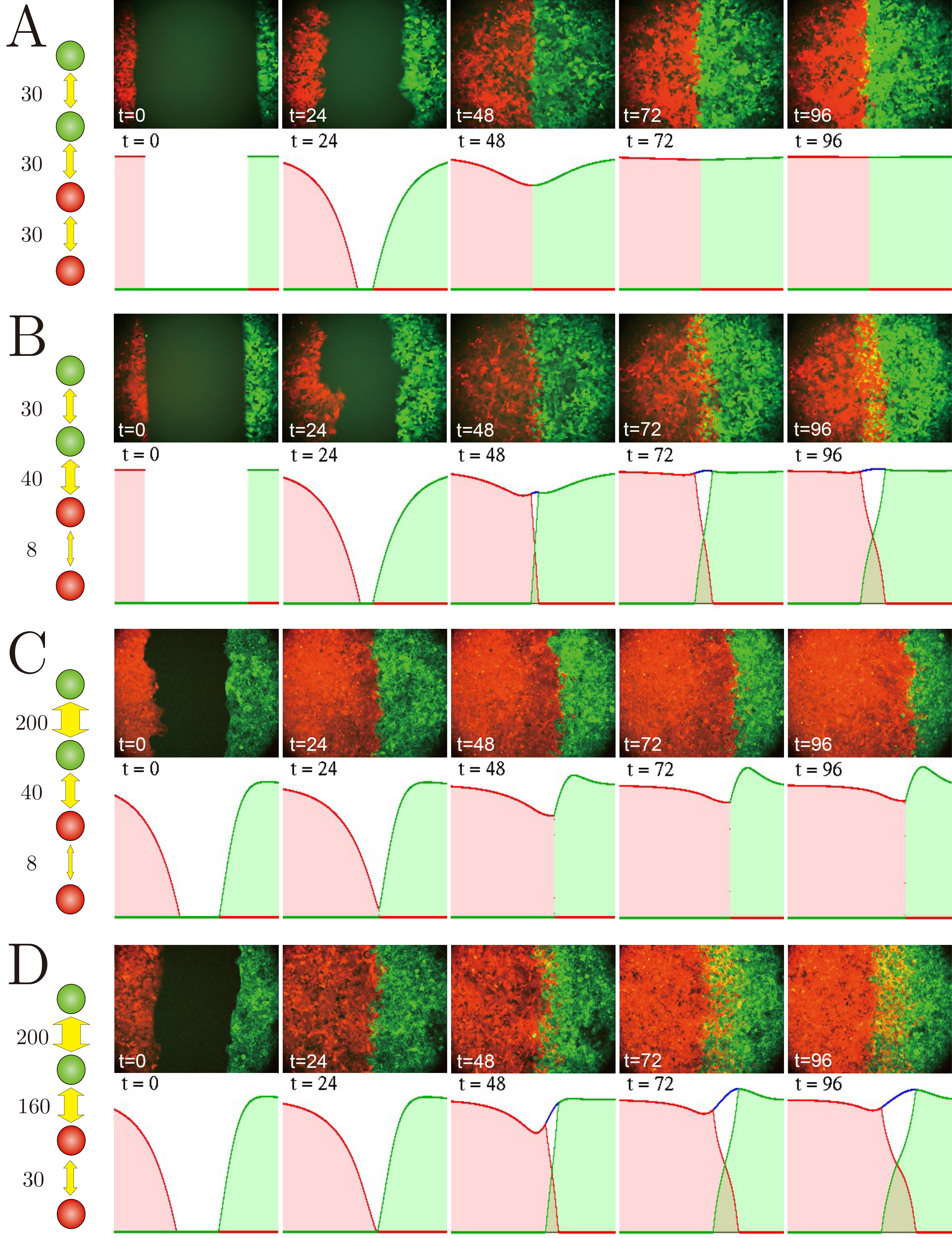}
\caption{
Experiments of co-cultures of HEK293 cells expressing different combination of nectins and cadherins, and the corresponding one-dimensional numerical simulations of \eqref{model_outside}: relative adhesion strengths and parameter values used in the numerical simulations (left),  experimental results (top) and numerical results (bottom). 
Time $t$ indicates elapsed time in hours.  
(A) Complete segregation: nectin-1 and N-cadherin (both green and red). 
(B) Bidirectionally invasion: nectin-1 and N-cadherin  (green); nectin-3 and N-cadherin (red). 
(C) Pushing back: nectin-3, N-cadherin and E-cadherin (green); nectin-3  and N-cadherin (red). 
(D) Asymmetric invasion: nectin-3, N-cadherin and E-cadherin (green); nectin-1  and N-cadherin (red). 
The experimental figures are from \cite{katsunuma} with permission from ROCKEFELLER UNIVERSITY PRESS. 
}
\label{fig_TE}
\end{figure}

Top figures of each Fig.~\ref{fig_TE}(A)--(D) show time-lapse microscopy of a mosaic forming assay of HEK293 cells expressing different nectins and cadherins. Depending on the combinations of cell-adhesion molecules, different patterns are observed. Note that  HEK293 cells naturally express N-cadherin, but not E-cadherin. Before the time-lapse experiments, these cells were separately cultured to allow the formation of independent colonies. When their colony edges came into contact with one another, their boundaries were examined. The shape variations of the colony edges at the start point are dependent on the experimental manipulation. 
Time $t$ indicates elapsed time in hours. 
In (A),  HEK293 cells expressing nectin-1, N-cadherin and EGFP (green) and those expressing nectin-1, N-cadherin and mCherry (red) were cultured. 
In early stage of culture, cells grow proliferously and expand their habitats. 
Each red cell moves from the left to the right and each green cell moves from the right to the left. And then, they fill in the gap around $t=30$h. 
They do not overlap at later times in this case. 
We can clearly see the boundary between green and red colonies. 
There is no driving force for intermingling with one another because the green and red cells are the same type. 
After the density reaches the carrying capacity, they stop growing and moving. 
In (B), HEK293 cells expressing nectin-1, N-cadherin and EGFP (green) and those expressing nectin-3, N-cadherin and mCherry (red) were cultured.  
In this case, they are intermingling with one another after contact, because each nectin-1+ green cell prefers to adhere to nectin-3+ red cells and vice versa. 
In (C), HEK293 cells expressing nectin-3, N-cadherin and E-cadherin-EGFP (green) and those expressing nectin-3, N-cadherin and mCherry (red) were cultured. 
These cells do not invade the counter colony each other. 
However, the red cells push back the counter colony of green cells. 
In (D), HEK293 cells expressing nectin-3, N-cadherin and E-cadherin-EGFP (green) and those expressing nectin-1, N-cadherin and mCherry (red) were cultured. 
The red cells invade the counter colony of green cells, but the green cells do not. The habitat (front position) of green cells does not change at later times.

We perform numerical simulations for these experiments in one space dimension using \eqref{model_outside}. 
The following actual measured values are used for the dimensional equations: $R=100\ \um$, $b_1=b_2=1/12\, \rhour^{-1}$, $k_1=k_2=0.0748\, \um^{-1}$. 
The sensing radius $R$ is determined by the length from the cell body to the tip of leading edge. 
The distance between the initial colonies is $500\, \um$. 
The dispersivity $c_\mathrm{p}$ controls the spreading speed of the colony edges. We set $c_\mathrm{p}=1.2\times 10^4\,  \um^3/\rhour$ such that the two colonies edges come into contact at 30 hours of incubation in Case (A). 
We could not measure the values $\phi$ and $m_i$, and simply chose $\phi=1\,\um^2/(\rhour \RU)$. 
It is thought that $m_i$ is greater than $k_i$ because the cells are moving due to cell adhesion even after the cells fill up the region. 
We chose $m_i=2k_i$. 
We consider the domain of length $2400\, \um$, that is, the computational domain is $(-L,L)$ with $L=12$. 
In order to compare numerical solutions with the experimental results, the visualized domain is set to be $[-4,4]$. 
The following Dirichlet boundary conditions are imposed: 
\begin{align*}
&\begin{array}{ll}
u(x,t)=0,\\
v(x,t)=k_2/m_2=0.5
\end{array}
\quad \mathrm{if}\quad x\le -L,\ t>0, \\
&\begin{array}{ll}
u(x,t)=k_1/m_1=0.5,\\
v(x,t)=0
\end{array}
\quad \mathrm{if}\quad x\ge L,\ t>0.
\end{align*}
We fix all parameters of the above for every simulations (A)--(D). 
The adhesive strength parameters are the only difference. 

We determined the adhesive strengths from the data given by Harrison et al.~\cite{hetal} and Katsunuma et al.~\cite{katsunuma}. 
Hereafter, we use ellipsis notation for the adhesive strength, e.g., $\mbox{n3NE-n1N}$ implies the adhesive strength between  nectin-3, N-cadherin and E-cadherin  expressing cells and nectin-1 and N-cadherin expressing cells.  
We chose $\mbox{n1N-n1N}=30\RU$(Response Unit) from Fig. 1a in \cite{hetal}, $\mbox{n3N-n3N}=8\RU$ from Fig. 1c in \cite{hetal}, and $\mbox{n1N-n3N}=40\RU$ from Fig. 1c in \cite{hetal}\footnote{We chose $\mbox{n1N-n3N}=230\RU$ in \cite{mt} using the datum from Fig. 1a in \cite{hetal}. But the datum ($=40\RU$) from Fig. 1c in \cite{hetal} is consistent with the experiment (B).}. 
Since $\mbox{n3NE-n3NE}$ is about 8 times of $\mbox{n2N-n2N}$ (Fig. 8E in \cite{katsunuma}), and $\mbox{n2N-n2N}=25\RU$ from Fig. 1b in \cite{hetal}, we chose $\mbox{n3NE-n3NE}=8\times 25\RU =200\RU$. 
There is no data available for $\mbox{n3NE-n3N}$ nor $\mbox{n3NE-n1N}$. 
Assuming that $\mbox{n3NE-n3N}$ (resp. $\mbox{n3NE-n1N}$) is similar to $\mbox{n2NE-n2N}$ (resp. $\mbox{n3NE-n2N}$), and using the data in Fig. 1b in \cite{hetal} and in Fig. 8E in \cite{katsunuma}, we chose $\mbox{n3NE-n3N}=40\RU$ and $\mbox{n3NE-n1N}=160\RU$. 
All parameters are summarized in Table~\ref{table_values}.

\begin{table}
\begin{center}
\begin{tabular}{|l|p{60mm}|p{25mm}|p{21mm}|}\hline
 & Description & Dimensional Value &
Dimensionless Value  \\\hline\hline

$t$ & Time & $1\, \rhour$ & 0.179520\\ \hline
$x$ & Spatial position & $1 \, \um$& 0.01 \\ \hline
$u(x,t), v(x,t)$ & Population densities at position $x$ and time $t$ & $\um^{-1}$& \\ \hline
$c_\mathrm{p}$ & Dispersivity & $1.2\times 10^4\, \um^{3}/\rhour$ & \\ \hline
$\phi$ & A constant of proportionality related to viscosity & $1\,\um^2/(\rhour \RU)$& \\ \hline
$R$ & Sensing radius & $100\, \um$& \\ \hline
$m_1=m_2$ & Crowding capacity & $2\times 0.0748\, \um^{-1}$& \\ \hline
$b_1= b_2$ & Proliferation rate & $1/12\, \rhour^{-1}$& 0.464201\\ \hline 
$k_1=k_2$ & Carrying capacity & $0.0748\, \um^{-1}$& 0.5\\ \hline
$L$ & Length of the computational domain $(-L,L)$& $1200\, \um$& $12$\\ \hline
 &  Distance between the initial colonies&$500\, \um$& $5$\\ \hline
$a_{ij}(i,j=1,2)$ & Adhesion strengths & & \\ 
 & \qquad $\mbox{n1N-n1N}$ & $30\RU$ & 0.25\\ 
 & \qquad $\mbox{n1N-n3N}$ & $40\RU$ & 0.333333\\ 
 & \qquad $\mbox{n3N-n3N}$ & $8\RU$ & 0.066667\\ 
 & \qquad $\mbox{n3NE-n3NE}$ & $200\RU$ & 1.666667\\ 
 & \qquad $\mbox{n3NE-n3N}$ & $40\RU$ & 0.333333\\ 
 & \qquad $\mbox{n3NE-n1N}$ & $160\RU$ & 1.333333\\ 
\hline\end{tabular}
\end{center}
\caption{Reference variables used in the numerical simulations.}
\label{table_values}
\end{table}

The numerical results of \eqref{model_outside} are shown in bottom figures of each Fig.~\ref{fig_TE}(A)--(D).  We use green line to plot $u$ and red line to plot $v$, and total population density $u+v$ is drawn with a blue line behind green and red lines. 
Appearance of blue line implies that the two types of cells are mixed. 
Under the green (resp. red) line is painted in translucent green (resp. pink) so that we can easily see whether the two populations are mixed or not. 
The step functions are used for the initial data in Cases (A) and (B). But, in the experiments in Cases (C) and (D), these cells are already moving at $t=0$, and the distributions at $t=0$ are similar to those in Cases (A) and (B) at $t=24$ (Watch Videos 3-6 in \cite{katsunuma} or Video~\ref{mov_TE}). 
Furthermore, the spreading speeds of the green colonies are slower than those of the red colonies. 
Therefore, 
the initial data of Cases (A) and (B) shifted a little to the left are set at $t=-24$ as the initial data in Cases (C) and (D), and we plot the numerical results from $t=0$. 
The numerical results demonstrate excellent agreement between the model~\eqref{model_outside} and the experiments, and illustrate that the model~\eqref{model_outside} is able to replicate each pattern not only qualitatively but also quantitatively. 
The interested readers can find Video~\ref{mov_TE} of the experiments versus the simulations using \eqref{model_outside} in the supplementary material. 

Katsunuma et al.~\cite{katsunuma} performed lattice-based numerical simulations in order to understand the asymmetric pattern formation in Fig.~\ref{fig_TE} (D). 
They noted ``our time-lapse observations revealed a significant difference in the migratory behavior of 293 cells, which was dependent of E-cadherin expression. ... The 293 cells expressing E-cadherin migrated more slowly toward the counter colony compared with those that did not express E-cadherin". 
Hence, they added cell mobility to their computational model  independently of cell adhesion, and concluded that differential mobility, in addition to adhesiveness, was responsible for the asymmetric mingling patterning 
even though there is a causal relation between cell adhesion and cell mobility as they mentioned. 
In contrast, Model~\eqref{model_outside} is capable of replicating the differential cell mobility as a consequence of the differential adhesion. 
The density of cells expressing E-cadherin near the propagating front is larger than that of cells which did not express E-cadherin. 
Model~\eqref{model_outside} can also replicate such phenomenon naturally. 

Finally, we turn our attention to the comparison of our modelling approach with respect to individual based modelling for cell adhesion. When single-cell-based models such as the cellular Potts models \cite{cgia,gg1992,potts,kapkgmh,man}, lattice-free models \cite{po,scp,HBTaylor} and the vertex dynamics models \cite{AGS,HNO,HOHK} are used, rules on adhesion, mobility, density, etc. have to be defined independently even if they have a causal relationship. And then, the number of parameters increases and the simulations become complicated and difficult to track and parameterize. Our model has the advantage of connecting the individual based modelling to macroscopic populations models as explained in the introduction with a minimal set of parameters amenable for easier tuning.

\section{Applications}

There are tight limitations in size and shape of the cells in single-cell-based models. 
However, the terminals of neurons show very irregular shapes that change dynamically.  
When investigating cell aggregations of such dynamic cells, it would be useful to use a continuous model. 
Here, we give brief summaries of two applications.

\subsection{Role of Reelin during layer formation in the mannmalian cerebral neocortex}
The mannmalian cerebral neocortex has a highly organized layered structure of neurons. 
During cortical development, a glycoprotein, Reelin, plays a crucial role in the neuronal migration and neocortical lamination because Reelin-deficient mice show disrupted neuronal organization. 
However, its precise role in neuronal layer organization remained unclear. 
It was thought that Reelin promotes neuronal adhesion and induces neuronal cell aggregation. 
To uncover how Reelin controls the intercellular adhesion among cortical cells, Matsunaga et al.~\cite{matsunaga} performed Reelin stimulation experiments. 
Two types of cells, radial glial cells (RGCs) and neurons (Neus), played key roles there. 
If Reelin promotes N-cadherin-dependent neuronal adhesion directly, the differential adhesivenesses should be in the order of Neu-Neu$>$Neu-RGC$>$RGC-RGC in the Reelin stimulation experiments. 
In this situation, according to the differential adhesion hypothesis, 
the neurons form clusters surrounded by the RGCs. 
However, the actual results were opposite, that is, the RGC clusters were engulfed by the neurons. 
A mathematical model based on \eqref{model_original} was used to understand the reason for this unexpected clustering pattern. 
The model predicted that Reelin causes an increase of cell-cell adhesion among neurons transiently but not persistently. 
Many expriments revealed that the prediction is correct. 
Finally, it was concluded that 
transient but not persistent increase in cell-cell adhesion might be necessary for the highly organized layered structure of neurons in the mammalian neocortex \cite{matsunaga}.

\subsection{Role of differential adhesion during columnar unit formation in the Drosophila brain}
\label{subsec_columnar}

In the developing visual center of the Drosophila brain, multiple neurons gather to form a columnar structure, a basic morphological and functional unit of the brain. Three types of core columnar neurons, R7, R8 and Mi1, play key roles at the initial step of column formation along a two dimensional layer of the brain during larval stage. A series of biological studies demonstrated that the differential adhesiveness of these neurons in the order of R7$>$R8$>$Mi1 causes their concentric arrangement. As a result, the terminal of R7 occupies the dot-like central region, the R8 terminal enwraps the R7 terminal forming a donut-like region and the Mi1 terminal occupies a grid-like region outside the R8 terminal. Since the neurites of columnar neurons show very irregular shapes that change dynamically, we formulated a three-component model based on \eqref{model_outside} by considering the density of neurites and differential adhesion between them. Our model demonstrated that the differential adhesion among R8, R7 and Mi1 is sufficient to reproduce the wild type and mutant patterns of the columns \cite{trush}.

\section{Further numerical experiments of the single-component model}

\subsection{Patterning under different initial data}

\begin{figure}[!ht]
\centering
\includegraphics[width=0.99\textwidth,clip]{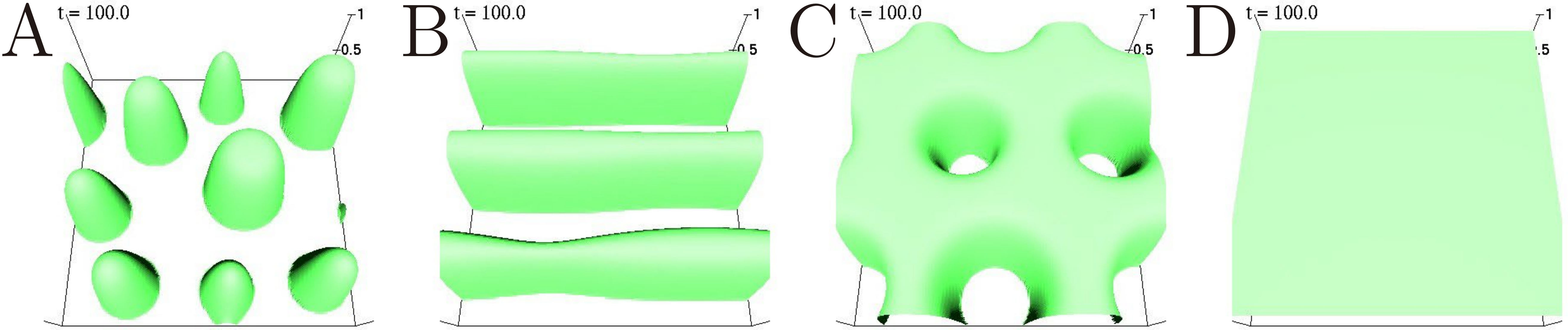}
\caption{
Numerical experiments of \eqref{eq_outside_1c} in two space dimensions with fixed $a=4$, $\omega\equiv 1$ and $f\equiv 0$, and different choices of initial data: (A) $V=0.2$, (B) $V=0.4$ , (C) $V=0.7$ and (D) $V=0.8$. 
}
\label{fig_patterns1}
\end{figure}

Cells form cell masses by adhesion, but the shape and size of the cell mass greatly differs according to the initial datum. 
Here, we show several different patterns depending on the difference of the initial data.
Numerical experiments of  Equation~\eqref{eq_outside_1c} are performed with fixed $a=4$, $\omega\equiv 1$ and $f\equiv 0$. 
The initial datum is set as a constant $V$ perturbed with 1\% random noise. 
Figure~\ref{fig_patterns1} shows that bell-shaped, striped and perforated patterns occur under variation of the initial cell density.  
The colony size and shape are different for each colony, and depend on the initial data. 

\subsection{Repulsive interaction and volume-dependent pattern formation}

\begin{figure}[!ht]
\centering
\includegraphics[width=0.99\textwidth,clip]{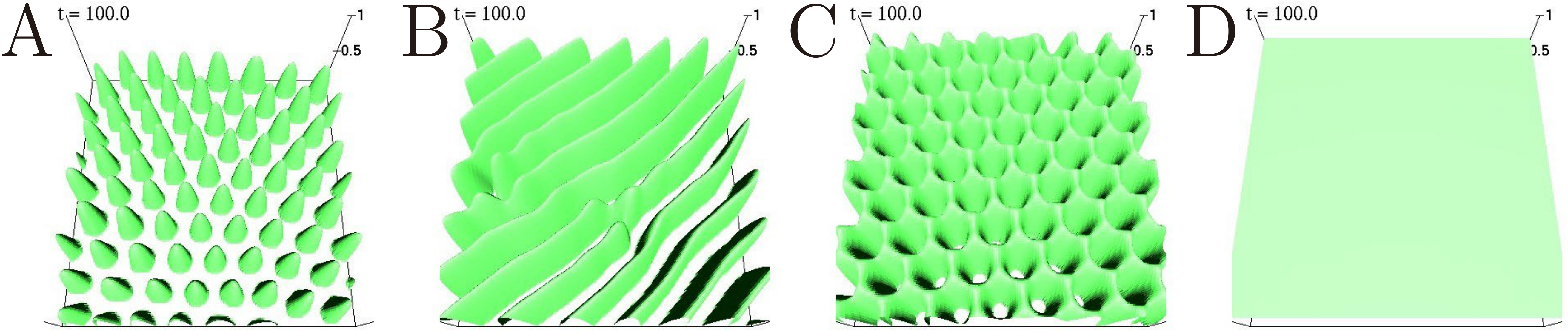}
\caption{
Numerical experiments of \eqref{eq_outside_1c} in two space dimensions with a short-range attractive and middle-range repulsive potential, and different choices of initial densities: (A) $V=0.1$, (B) $V=0.3$, (C) $V=0.5$ and (D) $V=0.7$. 
}
\label{fig_patterns2}
\end{figure}

\begin{figure}[!ht]
\centering
\includegraphics[width=0.5\textwidth,clip]{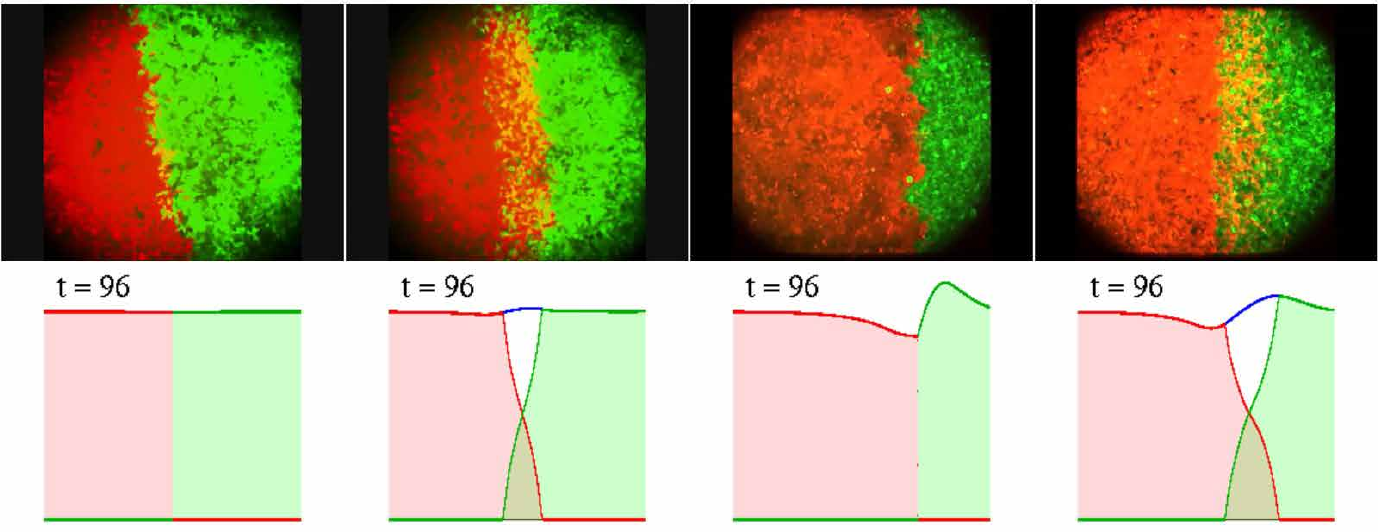}
\caption{
Movie of experiments of co-cultures of HEK293 cells expressing different combination of nectins and cadherins (top), and the corresponding one-dimensional numerical simulations of \eqref{model_outside} (bottom). The movie corresponds to Fig.~\ref{fig_TE}. 
The experimental movies are from \cite{katsunuma} with permission from ROCKEFELLER UNIVERSITY PRESS. 
}
\label{mov_TE}
\end{figure}

Cell sorting is occasionally caused not by cell-cell adhesion but by cell-cell repulsion, see, e.g., \cite{SuzukiSato,AMTaylor,HBTaylor} and references therein. 
The cell-cell repulsion is represented by our model just setting the adhesion parameter $a$ negative or setting $\omega$ negative. 
Here, recall that the function $\omega$  describes how the force is dependent on the distance from the cell. 
We carry out numerical simulations for Equation~\eqref{eq_outside_1c} with fixed $a=1$ and $f\equiv 0$. 
The function $\omega$ is set to be 
\begin{equation*}
\label{ar_omega}
\omega(r)=
\left\{
\begin{array}{lll}
a_1 & \mbox{if} & 0\le r < R_0, \\
a_2 & \mbox{if} & R_0\le r \le R=1,
\end{array}
\right.
\end{equation*}
where $0<R_0<1$ and $a_1$ and $a_2$ can  be positive or negative. 
For example, when $a_1$ is positive and $a_2$ is negative, it indicates short-range attractive and middle-range repulsive potential. 
We use the same initial datum as above. 
Figure~\ref{fig_patterns2} shows numerical results with fixed $R_0=0.85$, $a_1=4$ and $a_2=-32$, and different choices of $V$ that controls total mass. 
When the total mass is small, a spotted pattern arranged in order is observed (Fig.~\ref{fig_patterns2}A). 
The distance among the spots seems to be determined by $R_0$ and $R$. 
Here, we recall that the computational domain is a square with side length 10. 
When the total mass exceeds a certain threshold, a striped pattern appears (Fig.~\ref{fig_patterns2}B). 
When the total mass is increased further, a honeycomb pattern appears (Fig.~\ref{fig_patterns2}C). 
When the total mass is large enough, the density becomes homogeneous (Fig.~\ref{fig_patterns2}D). 
These patterns are robustly generated depending on the total mass,   
and distinct from the type of patterns in Fig.~\ref{fig_patterns1}. 

Thus, various patterns appear even with a single component. 
As mentioned in Section~\ref{subsec_columnar}, our model was applied to understanding the formation of columnar units in the Drosophila brain. 
Multiple columnar units are neatly arranged and form a hexagonal lattice just like Fig.~\ref{fig_patterns2}A. 
Our model might be able to explain the formation of such structures. 
Besides this, similar patterns to Fig.~\ref{fig_patterns2} appear in vegetation patterns~\cite{HMSZ} and patterns in animal skin~\cite{WK}. 
Our model has some advantages to elucidate such  pattern formations.

\section{Concluding remarks}

In this paper, we discussed and improved continuum models for cell-cell adhesion. Armstrong, Painter and Sherratt \cite{aps2006} proposed a celebrated model consisting of linear diffusions and nonlocal advection terms. However, it gives biologically unrealistic numerical solutions. In particular, it can not replicate mutually-immiscible phenomena. The underlying cause of the problem was that the model is based on random movement of each individual cell at the microscopic level. Murakawa and Togashi~\cite{mt} have changed the basic principles of cell movement from ``cells move randomly" to ``cells move from high to low pressure regions", and have proposed a modified continuous model for cell-cell adhesion based on the total population pressure. Their model was able to replicate the mutually-immiscible phenomena and the different types of sorting patterns. However, this model presents a drawback since it leads to corrugated surfaces of cell colonies which are not biologically reasonable.  In this paper, we proposed a modified model by rethinking of the nonlocal adhesion terms and imposing the saturation response to adhesive forces in a different manner. 
Numerical solutions of the modified model show smooth surfaces of cell colonies while being able to capture qualitatively the cell sorting mechanism. By a careful parameter choice, we are able to obtain an excellent agreement with the phenomena observed in experimental data by Katsunuma et al. \cite{katsunuma}. We also show how the models depends on the potential shape and we have explored not only cell-cell adhesion but also cell-cell repulsion.

\section*{Acknowledgments}
JAC was partially supported by the EPSRC grant number EP/P031587/1. 
HM was partially supported by JSPS KAKENHI Grant Numbers 26287025, 15H03635 and 17K05368, and JST CREST Grant No. JPMJCR14D3.  
HT was supported by JSPS KAKENHI Grant Numbers 18K06219, 18H04764, and a grant
from the Takeda Science Foundation.

\newpage 

\beginsupplement

\section*{Supplementary materials}

\end{document}